# Twisted polaritonic crystals in thin van der Waals slabs

Nathaniel Capote-Robayna[1], Olga Matveeva[2], Valentyn S. Volkov[2], Pablo Alonso-González[3,4,†] and Alexey Y. Nikitin[1,5,†].

[1]*Donostia International Physics Center (DIPC), Donostia-San Sebastián 20018, Spain.*

[2]*Center for Photonics and 2D Materials, Moscow Institute of Physics and Technology, Dolgoprudny, 141700, Russia.*

[3]*Department of Physics, University of Oviedo, Oviedo 3006, Spain.*

[4]*Center of Research on Nanomaterials and Nanotechnology, CINN (CSIC-Universidad de Oviedo), El entrego 33940, Spain.*

[5]*IKERBASQUE, Basque Foundation for Science, Bilbao 48013,Spain.*

*†Corresponding author. Email: alexey@dipc.org*

**Polaritons – hybrid light-mater excitations – are very appealing for the confinement of light at the nanoscale. Recently, different kinds of polaritons have been observed in thin slabs of van der Waals (vdW) materials, with particular interest focused on phonon polaritons (PhPs) – lattice vibrations coupled to electromagnetic fields in the mid-infrared spectral range with – in biaxial crystals, such as e.g. $\alpha$-MoO$_3$. In particular, hyperbolic PhPs – having hyperbola-like shape of their isofrequency curves – in $\alpha$-MoO$_3$ can exhibit ultra-high momenta and strongly directional in-plane propagation, promising novel applications in imaging, sensing or thermal management at the nanoscale and in a planar geometry. However, the excitation and manipulation of in-plane hyperbolic PhPs have not yet been well studied and understood. Here we propose a technological platform for the effective excitation and control of in-plane hyperbolic PhPs based on polaritonic crystals (PCs) – lattices formed by elements separated by distances comparable to the PhPs wavelength -, twisted with respect to the natural vdW crystal axes. In particular, we develop a general analytical theory valid for an arbitrary PC made in a thin biaxial slab. As a practical example, we consider a twisted PC formed by rectangular hole arrays made in MoO$_3$ slab and demonstrate the excitation of Bragg resonances tunable by the twisting angle. Our findings open novel avenues for both fundamental studies of PCs in vdW crystals**

**and the development of mid-infrared sensing and photodetection applications.**

In the last decade, polaritons in two-dimensional (2D) and van der Waals (vdW) materials (1, 2) have emerged as a promising mean to manipulate light on the nanoscale (3). Particularly, some natural polar uniaxial and biaxial crystals (such as e.g. h-BN (4), $MoO_3$ (5),$V_2O_5$ (6) or calcite (7)) support PhPs with either out-of-plane or in-plane directional propagation. In-plane anisotropic PhPs are particularly attractive because they manifest intriguing optical phenomena, such as light canalization (8), topological transitions (9) or negative refraction (10), among others, which are directly accessible in real space via near-field nanoimaging techniques.

PCs based on anisotropic polar vdW materials (such as h-BN) have demonstrated interesting possibilities for enhanced light-matter interactions (11-13) and topological photonics (14). On the other hand, PCs in in-plane anisotropic materials, where the manipulation of polaritons can exhibit more degrees of freedom, have not yet been addressed.

Here we introduce the concept of PCs created in slabs of biaxial crystals (supporting in-plane anisotropic polaritons), with the PCs lattice vectors twisted with respect to the crystal optical axes. We demonstrate that the polaritonic Bragg resonances emerging in these PCs can be efficiently tuned, offering a new way of controlling light in deeply subwavelength scales. We develop a simple theoretical approach for treating the diffraction of light by such twisted PCs and prove its validity for realistic structures (e.g. hole arrays) by comparing with full-wave electromagnetic simulations.

We first briefly outline the theoretical technique we have developed in this work. It has some important approximations, which allow one to significantly simplify the mathematical treatment of the diffraction by an arbitrary lattice made in a thin biaxial crystal slab. Specifically, let us consider a double-periodic structure such as, e.g., repetition of holes with periods $L_1$ and $L_2$ along two perpendicular directions, made in a thin biaxial slab (Figure 1a). The slab is sandwiched between two semi-infinite dielectric media with dielectric permittivities $\varepsilon_1$ and $\varepsilon_2$. For sufficiently thin slabs (with the thickness much smaller than the polariton wavelength), we can avoid considering the fields inside the slab, thus replacing it in theoretical formalism by a two-dimensional (2D) in-plane anisotropic conducting sheet placed in the $xy$-plane (see Supporting Information, section S1) with an effective 2D conductivity tensor $\hat{\sigma} = \frac{cd}{2i\lambda_0}\hat{\varepsilon}$ (15) (for convenience, we will use in the following its normalized value, $\hat{\alpha} = \frac{2\pi}{c}\hat{\sigma}$). In this formalism it is sufficient to properly match the fields on the effective conducting sheet by the boundary conditions, taking into account the continuity (discontinuity) of the in-plane electric (magnetic) fields. The introduction of the effective 2D conductivity tensor is valid for the description of the lowest electromagnetic waveguiding mode in the slab (the one possessing the longest wavelength). Then, in general, we assume that the effective normalized conductivity tensor of the slab is a periodic function of the coordinates, $\hat{\alpha}(\mathbf{r}) = \hat{\alpha}(\mathbf{r} + n_1\mathbf{L}_1 + n_2\mathbf{L}_2)$, with $n_1, n_2$ being integers, so that it can be expanded into the Fourier series:

$$\hat{\alpha}(\mathbf{r}) = \sum_N \hat{\alpha}_N e^{i\mathbf{G}_N \mathbf{r}}, \quad (1)$$

where capital letter $N$ is a compacted multi-index meaning both $n_1$ and $n_2$ indices ($N \equiv \{n_1, n_2\}$); and $\mathbf{G}_N$ is a general reciprocal vector $\mathbf{G}_N = n_1\mathbf{g}_1 + n_2\mathbf{g}_2$. For compactness, we will use Dirac notations, in which the in-plane components of the s- and p- polarization basis vectors for each multi-index $N$ read as:

$$|sN\rangle = \frac{1}{k_{tN}}\begin{pmatrix}-k_{yN}\\k_{xN}\end{pmatrix}e^{i(xk_{xN}+yk_{yN})} \quad |pN\rangle = \frac{1}{k_{tN}}\begin{pmatrix}k_{xN}\\k_{yN}\end{pmatrix}e^{i(xk_{xN}+yk_{yN})}, \quad (2)$$

where $\mathbf{k}_{tN} = (k_{xN}, k_{yN})$ are the in-plane momenta, and $k_{tN}$ stand for their norm. Then, the in-plane components of the electric fields in the superstrate (medium 1, $z > 0$) and in the substrate (medium 2, $z < 0$) can be represented as a Fourier-Floquet series constituting the superposition of plane waves of both s- and p- polarizations:

$$\mathbf{E}_{1t}(z) = (I_s|s0\rangle + I_p|p0\rangle)e^{-izk_{inc}}$$
$$+ \sum_N [(R_{sN}|sN\rangle + R_{pN}|pN\rangle)e^{-izk_{z1N}}] \qquad z \geq 0, \qquad (3)$$

$$\mathbf{E}_{2t}(z) = \sum_N [(T_{sN}|sN\rangle + T_{pN}|pN\rangle)e^{izk_{z2N}}] \qquad z < 0, \qquad (4)$$

where $I_{s,p}$, stand for the amplitudes of the incident field, $R_{s,pN}$ and $T_{s,pN}$ present the amplitudes of the scattered (diffracted) plane waves with the indices s and p indicating their polarizations. The out-of-plane momentum is given by $k_{ziN} = \sqrt{\varepsilon_i k_0^2 - k_{tN}^2}$, where $i = \{1,2\}$ labels the medium, and $k_{inc} = \sqrt{\varepsilon_1 k_0^2 - k_{t00}^2}$.

From Eqs 3, 4 we can find similar expressions for the magnetic fields, with the help of Maxwell's equations (see Supporting Information, section S1). Using the following boundary conditions

$$\mathbf{E}_{1t}(z = 0) = \mathbf{E}_{2t}(z = 0), \qquad (5)$$

$$\mathbf{e}_z \times (\mathbf{H}_{1t}(z = 0) - \mathbf{H}_{2t}(z = 0)) = 2\hat{\alpha}\mathbf{E}_{1t}(z = 0), \qquad (6)$$

and after some straightforward algebra (see Supporting Information, section S1), we arrive at a linear system of algebraic equations for the amplitudes of the scattered plane waves, which can be compactly written as:

$$\sum_{\beta N} D_{NN'}^{\beta\beta'} T_{\beta N} = 2I_{\beta'} Y_{inc}^{\beta'} \delta_{00N'}, \qquad (7)$$

$$D_{NN'}^{\beta\beta'} = (Y_{1N}^{\beta} + Y_{2N}^{\beta})\delta_{\beta\beta'}\delta_{NN'} + 2M_{NN'}^{\beta\beta'}, \qquad (8)$$

where for each polarization we defined $Y_{iN}^s = k_{ziN}/k_0$ and $Y_{iN}^p = \varepsilon_i k_0/k_{ziN}$, and $Y_{inc}^{\beta'}$ takes values $Y_{inc}^s = k_{inc}/k_0$ and $Y_{inc}^p = \varepsilon_1 k_0/k_{inc}$, for s- and p-polarization, respectively. The matrix elements $M_{NN'}^{\beta\beta'}$ are composed by the products between the conductivity tensor and the in-plane momentum:

$$M_{NN'}^{\beta\beta'} = \frac{1}{k_{tN}k_{tN'}} \Big[ (-1)^{1+\delta_{\beta\beta'}} \alpha_{N'-N}^{xx} k_{\bar{\gamma}N} k_{\bar{\gamma}'N'}$$

$$+ \alpha_{N'-N}^{xy} \left( (-1)^{1+\delta_{\beta p}} k_{\gamma'N} k_{\bar{\gamma}N'} + (-1)^{1+\delta_{p\beta'}} k_{\bar{\gamma}'N} k_{\gamma N'} \right) \quad (9)$$

$$+ \alpha_{N'-N}^{yy} k_{\gamma N} k_{\gamma'N'} \Big],$$

where $\gamma$ stays for $x$ or $y$ when $\beta$ takes s or p values, respectively, while oppositely, $\bar{\gamma}$ stays for $x$ or $y$ when $\beta$ takes p or s values, respectively. Note that $\gamma$ is a function of $\beta$, while $\gamma'$ depends upon $\beta'$. By truncating the infinite system 7 to a maximal order, $N_{max}$ (sufficiently large to achieve convergence, see Supporting Information, section S4), one can straightforwardly invert the matrix $\widehat{D}$ by standard numerical procedures and obtain all the unknown amplitudes of the diffracted waves. Then, both the reflection and transmission coefficients can be calculated, as well as the field distribution above the structure. Additionally, as we demonstrate below, the system of equations 7 can also be solved analytically in the tight-binding approximation. Independently upon the solution method, it is important to note that, as it follows from Eq 9, the non-diagonal elements of the matrix $\widehat{D}$, describing the interaction between the diffracted plane waves, are proportional to the Fourier harmonics of the normalized conductivity. Therefore, the Fourier decomposition of the periodic structure plays a crucial role in the diffraction of an incident plane wave and resonant excitation of polaritonic modes.

Figure 1a shows an example of a square hole array (HA), with its Fourier coefficients represented in Figure 1b. The lattice vector $L_1$ forms an angle $\phi$ (twisting angle) with respect to the [100] in-plane crystallographic axis. The HA can be described by the spatially-dependent normalized conductivity, $\hat{\alpha}(r) = \hat{\alpha}_0(1 - f(r))$, where the step function $f(r)$ takes a value of 1 (0) inside (outside of) the holes with radius $a$, respectively. As typical for such periodic functions, the Fourier transform (FT) coefficient $\hat{\alpha}_{00}$ takes the highest value, while the amplitudes of the other Fourier harmonics decrease with $N$. Without loss of generality and for illustrative purposes, we will focus in this paper on the excitation of polaritons in a periodically structured α-MoO₃ slab.

The vdW crystal α-MoO$_3$ has several optical phonons at mid-IR frequencies, which open frequency bands (Reststrahlen bands defined between the longitudinal (LO) and transversal (TO) optical phonons frequencies) that support the propagation of in-plane anisotropic PhPs (16). These PhPs form a set of in-plane anisotropic waveguiding modes in thin slabs, typically named as $Mn$, with $n = 0,1, ...$ etc, with $n$ meaning the quantization of the mode in the transversal (perpendicular to the slab) direction (17, 18). The wavelength of the modes as well as their propagation length decreases with $n$, so that the $M0$ mode exhibits the longest polaritonic wavelength and can be treated safely within our theoretical approximation.

We will focus on polaritonic effects originating from the first-order scattering processes of the polaritonic M0 mode, namely the processes provided by the first-order Fourier harmonics of the 2D lattice. For this reason, we will investigate an artificial lattice composed by exclusively first-order Fourier coefficients of $\hat{\alpha}$ (a "harmonic" lattice), i.e. with the orders $(0,0)$, $(\pm 1,0)$ and $(0, \pm 1)$. On the other hand, we will take the amplitudes of these Fourier harmonics from the realistic expansion of the HA (marked by the dashed white circles in Figure 1b). Furthermore, let us first assume that the lattice Fourier harmonics along the *y*-direction have zero amplitude, $\hat{\alpha}_{0\pm 1} = \hat{0}$, so that the lattice is one-dimensional (1D). Its schematics is shown in Figure 2a. The 1D lattice is aligned along the [100] crystal direction (coinciding with the *x*-axis), and the polarization of the normally-incident light is parallel to the *x*-axis, thus being aligned with the grating vector, $\boldsymbol{G}$. Although our theoretical approach is valid for arbitrary isotropic dielectrics surrounding the slab, we assume that both dielectric media have a dielectric permittivity equal to 1. On the one hand, this choice simplifies the interpretation and analysis of the resonances obtained, and, on the other hand, it mimics the typical experimental scheme in which the α-MoO$_3$ slab is placed on a highly transparent substrate (e.g. BaF$_2$) (16) . Note that as soon as the lattice cannot provide any momenta in the *y*-direction (and the incident wave is polarized along the *x*-direction), the k-vectors of all the diffracted waves, as well as their electric fields, belong to the *xz*-plane. Since for normal incidence the incident plane is not well-defined, we assign the p-polarization to the plane waves having their electric fields within the *xz*-plane, while the "s-polarized" waves are completely absent.

Figure 2c shows the spectra of the normalized reflection coefficient (introduced as $\delta R = \left|\frac{R_{p00} - R^b_{p00}}{R^b_{p00}}\right|$, where $R^b_{p00}$ stands for the p-polarization reflection coefficient of a bare slab) and the amplitudes of the first, $(\pm 1, 0)$, and second order, $(\pm 2, 0)$, field harmonics. Due to the symmetry provided by the normally incident wave, the amplitudes of the field harmonics $(n, 0)$ and $(-n, 0)$ are identical. The parameters of the lattice (with period $L = 900\ nm$ and hole radius, $a = 200\ nm$) have been chosen to guarantee the emergence of PhP resonances within the second hyperbolic Reststrahlen band, RB2, of α-MoO₃ (16). i.e., in the frequency range $840 - 960\ cm^{-1}$. At a frequency $\omega = 909.9\ cm^{-1}$ all the field amplitudes show a prominent and narrow resonant peak (with a quality factor $Q$ for $|T_{p10}|^2$ of ~ 150), with the largest value reached by the first-order field harmonic, thus indicating that the latter has the major contribution to the resonance. Besides, the position of the peak matches with the condition of the first-order Bragg resonance, $\boldsymbol{k}_{t10} = \boldsymbol{G}$, as seen from comparison with both the hyperbolic isofrequency curve (IFC) –a slice of the three-dimensional dispersion surface at a constant frequency– of the M0 mode in α-MoO₃ slab (Figure 2b) and its dispersion (Figure 2d). Finally, the distribution of the electric field above the lattice (Figure 2a), reconstructed with the help of Eq 3, visualizes a standing wave with oscillation length (distance between maxima with the same polarity – blue or red in the figure) matching the period of the lattice. These observations clearly prove that the spectra in Figure 2c manifest the excitation of a narrow first-order Bragg PhP resonance on a lattice realized in the α-MoO₃ slab.

Remarkably, the numeric solution of the linear system 7 (continuous curves in Figure 2c) shows an excellent agreement with our analytical approximation (square symbols in Figure 2c), in which only zero-order and first-order field harmonics are retained (see Supporting Information, section S3). This agreement demonstrates the validity of the perturbational approach (analogous to(15, 19)), in which retention of only a few field Fourier harmonics is sufficient.

Let us now study the dependence of the emerging Bragg PhP resonances upon the twisting angle, $\phi$ (angle between the grating vector, $\boldsymbol{G}$, and the [100] direction of the α-MoO₃ crystal slab) in case of the simplest 1D harmonic lattice (see schematics in Figure 3a). As previously, we assume an illumination of the structure by a normally incident wave with the electric field polarized along the *x* axis, the latter aligned with the α-MoO$_3$ [100] crystal direction. Figure 3a shows the numerically and analytically calculated $\delta R$ as a function of frequency (solid curves and square symbols, respectively) for different values of $\phi$. For $\phi \neq 0$, the polarization conversion takes place so that the cross-polarization (s-) components of the diffracted plane waves appear. The amplitude of the s-polarized reflected wave in the zero diffraction order is represented in Figure 3a by dashed lines, indicating a polarization conversion in the order of 5%. More importantly, the resonant peak associated with the first-order PhP Bragg resonance strongly redshifts with increasing $\phi$. The redshift of the resonance can be understood by the Bragg resonance condition, i.e. matching of the $\boldsymbol{k}_{t\pm10}$ in-plane wavevectors with the IFC of the M0 mode (see Figure 3b). Indeed, due to the rotation of the hyperbolic IFC with increasing $\phi$, the IFC meets the points (1,0) and (-1,0) in the reciprocal space at a lower frequency (the asymptote of the hyperbola becomes more "aligned" along the $k_x/k_0$), the latter perfectly matching the position of the resonant peaks in Figure 3a. Obviously, for the extreme case of $\phi = 90°$, the hyperbolic IFC (black curve in Figure 3b) does not meet the reciprocal lattice vector for any frequency and thus the PhP Bragg resonance does not appear ($\delta R$ coefficient is close to zero). Taken together, these results show, for the first time, the tuneability of the PhP Bragg resonance by simply twisting the photonic lattice with respect to the vdW crystal slab. In practice, such tuning can be realized for lattices made both directly in a biaxial vdW crystal slab and in a periodically-structured substrate with a continuous rotatable crystal slab placed on top of it.

A much richer family of rotation-tunable Bragg resonance appears when the periodicity in the biaxial crystal slab is simultaneously formed in two crystal directions. An example of such 2D lattice (consisting of the two simplest 1D harmonic lattices with mutually perpendicular Bragg vectors and twisted at an angle $\phi = 60°$ with respect to the α-MoO$_3$ [100] crystallographic axis) is represented as a schematic in Figure 4a. Illuminating the periodic structure by a normally-incident plane wave polarized along the α-MoO$_3$ [100] crystallographic axis leads to the emergence of multiple resonant peaks in the spectra of $\delta R$, as shown in Figure 4b (red curve and square symbols for the numerical and analytical solutions, respectively). We assume that the two most prominent peaks appearing at $860\ cm^{-1}$ and $895.6\ cm^{-1}$ can be associated with the (±1,0) and (0,±1) first-order Bragg resonances, respectively, as can be concluded from matching the reciprocal lattice vectors $\boldsymbol{G}_{10}$ and $\boldsymbol{G}_{01}$ with the hyperbolic IFC in the two orthogonal directions (Figure 4c,d). Remarkably, the transmission coefficient $T_{p10}$ (Figure 4b, black curve) has a single maximum at $860\ cm^{-1}$, while the coefficient $T_{p01}$ (Figure 4b, green curve) has a single maximum at $895.6\ cm^{-1}$. Since these transmission coefficients characterize the amplitudes of the first-order diffracted waves (±1,0) and (0,±1), their resonance character indicates the formation of intense polaritonic standing waves (Bloch waves) along the *x* and *y* directions. The latter can be clearly seen in the calculated field distribution shown in Figure 4a, confirming our assumption regarding the nature of the resonances responsible for the peaks of $\delta R$. Notice that while the amplitudes of the (±1,0) and (0,±1) diffraction orders are dominating in the whole Fourier-Floquet expansion given by Eqs. 3, 4 (particularly, at the resonance frequencies, as expected), the amplitudes of the higher diffraction orders quickly decay with *N* (see for instance the amplitude of the second order, shown by the magenta curve in Figure 4a). In fact, such behavior of the diffraction amplitudes is expected from the Fourier composition of our diffraction grating. The latter mainly generates the first-order diffracted waves via the first-order scattering processes (proportional to $\hat{\alpha}_{0\pm1}$), while the higher-order diffracted waves are generated by the higher-order scattering processes (proportional to higher powers of $\hat{\alpha}_{0\pm1}$ and thus resulting in much smaller diffraction amplitudes, particularly for rather "weak" gratings) (15, 19). Non-collinearity of the incident wave polarization with the lattice vectors (analogously to the case of the twisted 1D grating) enables the transformation of the polarization, as illustrated by the spectra of the amplitude of the s-polarization component of the zero-order reflected wave, $R_{s00}$ (magenta curve in Figure 4). Although the efficiency of this transformation is small, it can be substantially enhanced by properly fitting the parameters of the grating as well as its orientation with respect to the crystal axes and incident field. Altogether, our analysis of PhP Bragg resonances in the 2D lattice in α-MoO₃ slab, twisted with respect to the crystal axes, suggests interesting possibilities for the versatile control of PhPs by means of the twisting angle or the lattice amplitude and periods. In particular, twisting the lattice with respect to the crystal axes can enable the excitation of PhPs in predetermined directions. Namely,

PhPs can be either excited in different directions at the same frequency, or in different directions at different frequencies, both scenarios being attractive for potential electromagnetic routing applications in the near field.

Finally, we apply our theory to illustrate PhP Bragg resonances in an experimentally-realizable structure consisting of a periodic HA in an α-MoO$_3$ slab, illustrated in Figure 1a. Such HA can be realized, for instance, by decorating the slab with periodically-arranged circular holes milled by high-resolution focused ion beam (FIB) lithography (20). Figure 5a shows the spectra for the reflection, transmission and absorption coefficients (ρ, $\tau$ and A, respectively) of normally-incident light when considering a square HA with periodicity $L = 900\ nm$ and hole radius, $a = 200\ nm$ (the unit cell together with the polarization of the incident light and axes orientation is sketched in the inset to Figure 5a). In all the spectra shown, a number of other resonant features appear at different frequencies. Obviously, as all the spatial Fourier harmonics of the HA have nonzero value, its optical response has a significantly more complex structure, as compared to the harmonic lattices considered above. For the interpretation of such emerging resonances, in Figure 5b we represent the calculated IFCs of the M0 PhP mode in the reciprocal space at the resonant frequencies. By monitoring the crossings between the IFCs and points in the reciprocal space, we assign the specific Bragg resonances (of certain orders) to the corresponding peaks/dips in the spectra in Figure 5a. Notice that the "visibility" of the resonance (relative difference between the maximal and minimal values of the reflectivity/absorption within the resonance) decreases with the resonance diffraction order, consistently with the decrease of the lattice Fourier harmonic amplitudes (Figure 1b) at higher orders, resulting in a weaker resonant scattering processes. Remarkably, the strongest resonant feature in Figure 5a matches with the Bragg condition for the (±1,±2) resonance, corresponding to the nearly asymptotic regime of the hyperbolic IFC (a straight line), where the PhP momentum is already rather high. We speculate that apart from the Bragg scattering processes, the high efficiency of the resonant PhP excitations in (±1,±2) diffraction orders can be related with the dipolar resonance of the holes, as well as with the flat bands arising in polaritonic crystals (11). However, the detailed analysis of such anomaly goes beyond the scope of this manuscript and will be addressed in future works.

To summarize, we have illustrated rotation-tunable anisotropic PhP Bragg resonances in twisted lattices formed in biaxial vdW crystal slabs. We have developed a simple theoretical approach suitable for both numerical and analytical analysis of "collective" polaritonic resonances in arbitrary lattices, which is valid for any biaxial crystal slab, provided its thickness is smaller than the wavelength of polaritons. We have clarified the role of the spectral Fourier composition of the lattices, considering their specific Fourier harmonics. Our results open a new research area of twisted polaritonic crystals, including those supporting non-trivial topological polaritons. These crystals can be very appealing for some practical use, as e.g. in tunable sensors or photodetectors.

## METHODS

**Numerical simulations.** Full-wave simulations, based on the finite-element method (FEM) in the frequency domain, were performed using COMSOL. We simulated the two-dimensional square periodic hole array in $MoO_3$ slab with thickness $d = 100\ nm$, period $L = 900\ nm$, and hole radius $a = 200\ nm$. The incident angle of an illuminating p-polarized plane wave, as well as the angle between one of the Bragg vectors of the hole array and [100] crystal axis were varied (see Supporting Information, section S5). The dielectric permittivity of the superstrate was taken for the air $\varepsilon_1 = 1$, while $\varepsilon_2$ (substrate) was assumed to be that of $BaF_2$.

## Supporting Information

Analytical derivation of the general system of equations for the amplitudes of the scattered fields; Simplified solution for the harmonic lattices; Convergence of FT harmonics and transmittance coefficient; Comparison between the analytical solution and full-wave simulations for a hole array in α-$MoO_3$ slab.

## ACKNOWLEDGMENTS

A.Y.N. acknowledges the the Spanish Ministry of Science and Innovation (grant PID2020-115221GB-C42) and the Basque Department of Education (grant PIBA-2020-1-0014). P.A.-G. acknowledges support from the European Research Council under starting grant no. 715496, 2DNANOOPTICA and the Spanish Ministry of Science and Innovation (State Plan for Scientific and Technical Research and Innovation grant number PID2019-111156GB-I00). O.M. and V.S.V. acknowledge the

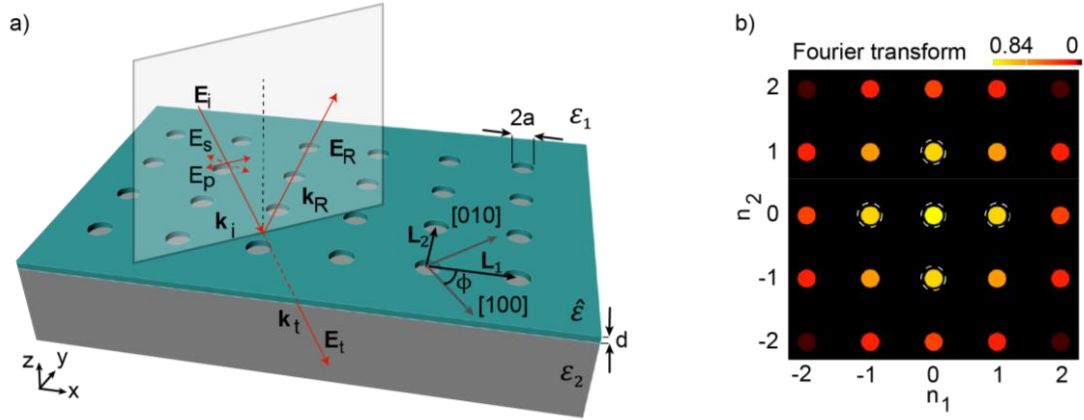

**Figure 1. An example of a twisted PC. (a)** Schematics of a square HA in a biaxial slab with thickness d, permittivity tensor $\hat{\varepsilon}$, and circular holes with radius $a$. The HA is defined by its vector basis $L_1$ and $L_2$ (in the case shown their lengths are equal $L_1 = L_2 = L$). The HA is twisted at an angle $\phi$ with respect to the crystallographic axes in the biaxial slab. The HA is illuminated by an incident plane wave. **(b)** Absolute value of the Fourier coefficients of the periodic HA (colorplot). The white dashed circles highlight the Fourier harmonics considered in the studied harmonic lattices.

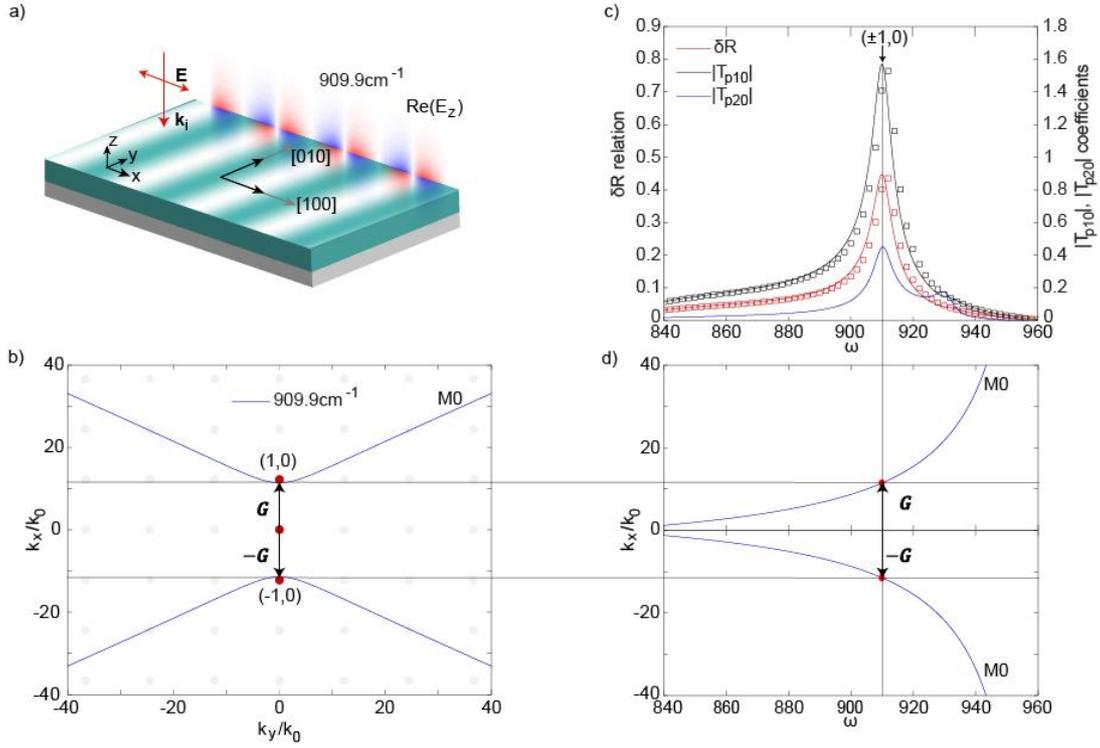

**Figure 2. PhP Bragg resonance in a 1D harmonic lattice in an α-MoO₃ slab. (a)** Schematics of the 1D lattice with its vector aligned to the α-MoO₃ [100] crystallographic axis. The lattice period is $L = 900\ nm$, the permittivities of the surrounding dielectric media are $\varepsilon_1 = \varepsilon_2 = 1$ and the MoO₃ slab thickness is $d = 100\ nm$. The amplitude of the lattice Fourier harmonic $\hat{\alpha}_{\pm 10}$ is taken for a HA with hole radius $a = 200\ nm$. The normally-incident light is polarized along the *x*-axis. The vertical electric field distribution shown is calculated at $\omega = 909.9\ cm^{-1}$. **(b)** Isofrequency curve (IFC) of the M0 PhP mode at $\omega = 909.9\ cm^{-1}$. $\boldsymbol{G} = \pm\boldsymbol{g}_{10}$ represents the reciprocal lattice vector. The reciprocal lattice points corresponding to the lattice Fourier coefficients with non-zero values are highlighted in red. **(c)** Spectra of the amplitudes of the (±1,0) and (±2,0) diffracted waves ($T_{p10}$ and $T_{p20}$, respectively) and the relative reflection coefficient, $\delta R$ **(d)** Dispersion relation for the M0 PhP mode.

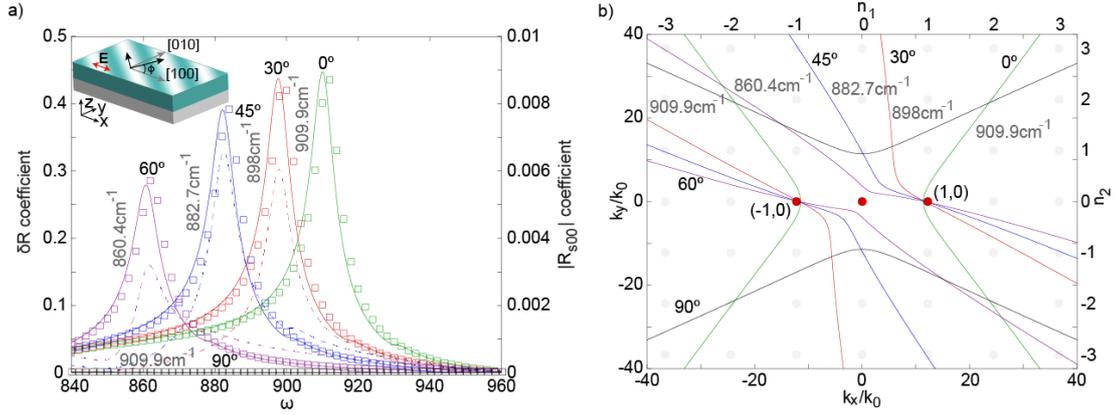

**Figure 3. PhP Bragg resonance in a twisted 1D harmonic lattice in a MoO₃ slab.** **(a)** Numerically and analytically calculated spectra of $\delta R$ (shown by continuous curves and square symbols, respectively) for different values of the twisting angle $\phi$. The discontinuous lines trace $R_{s00}$ coefficient for the same angles. The schematics of a 1D grating with its basis twisted an angle $\phi$ with respect to the crystallographic axes is shown as an inset. Normal incidence light with linearly polarized electric field along x-axis is considered. The parameters of the structure are: $L = 900\ nm$, $\varepsilon_1 = \varepsilon_2 = 1$, $d = 100\ nm$. The $\hat{\alpha}_{\pm 10}$ amplitude is taken for the HA with $a = 200\ nm$. **(b)** IFCs for fixed frequencies and $\phi$. The reciprocal space points with the corresponding non-zero Fourier coefficients of the conductivity tensor are highlighted in red.

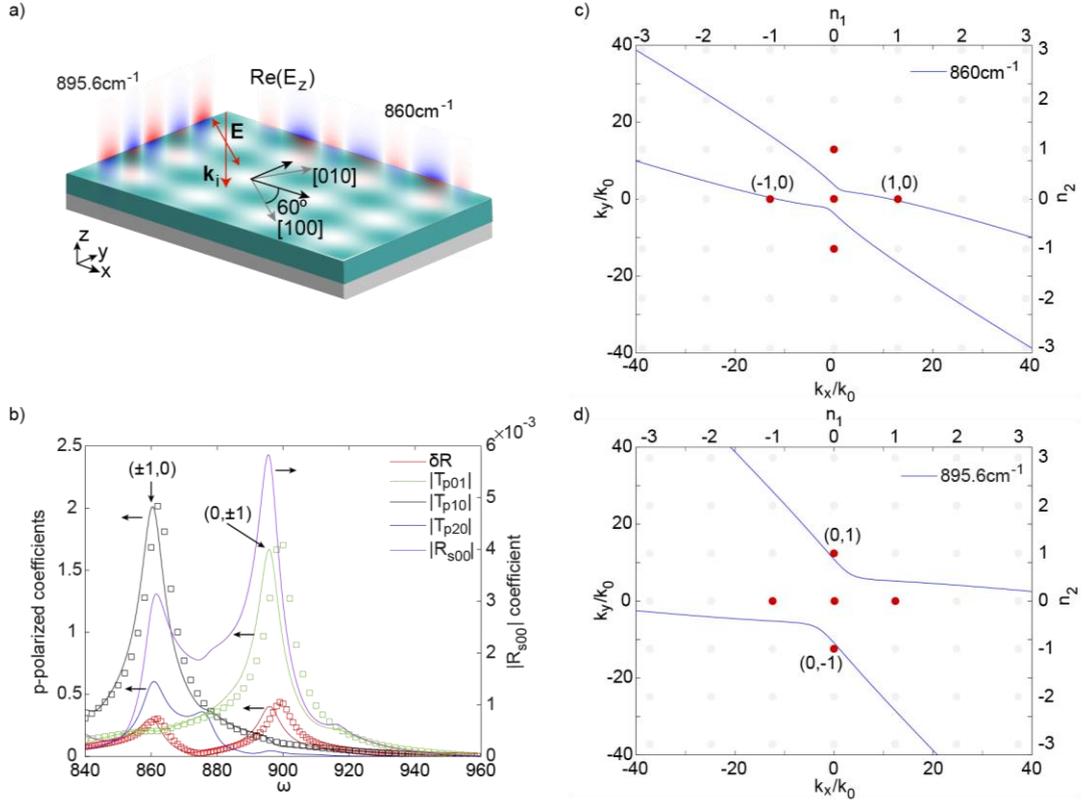

**Figure 4. PhP Bragg resonance in a twisted 2D harmonic lattice in an α-MoO₃ slab.** (a) Schematics of the 2D lattice twisted by $\phi = 60°$ with respect to the α-MoO₃ [100] crystallographic axis. The lattice period is $L = 900\ nm$, the permittivities of the surrounding dielectric media are $\varepsilon_1 = \varepsilon_2 = 1$ and the MoO₃ slab thickness is $d = 100\ nm$. The amplitude of both $\hat{\alpha}_{\pm 10}$ and $\hat{\alpha}_{0\pm 1}$ lattice Fourier harmonics are taken for a HA with $a = 200\ nm$. The normally-incident light is polarized along the α-MoO₃ [100] crystallographic axis. The vertical electric field spatial distribution shown is calculated at $\omega = 860\ cm^{-1}$ and $\omega = 895.6 cm^{-1}$, revealing polaritonic Bloch waves oscillating along the x and y-axis, respectively. (b) Spectra of the amplitudes of the (±1,0), (0,±1) and (±2,0) diffracted waves ($T_{p10}$, $T_{p01}$ and $T_{p20}$, respectively), the relative reflection coefficient, $\delta R$, and the cross-polarization $R_{s00}$ coefficient. The results of the single-harmonic approximation are represented by the square symbols. (c,d) IFCs for frequencies $860\ cm^{-1}$ and $895.6 cm^{-1}$, and twisting angle $\phi = 60°$. The reciprocal space points with the corresponding non-zero Fourier coefficients of the conductivity tensor are highlighted in red.

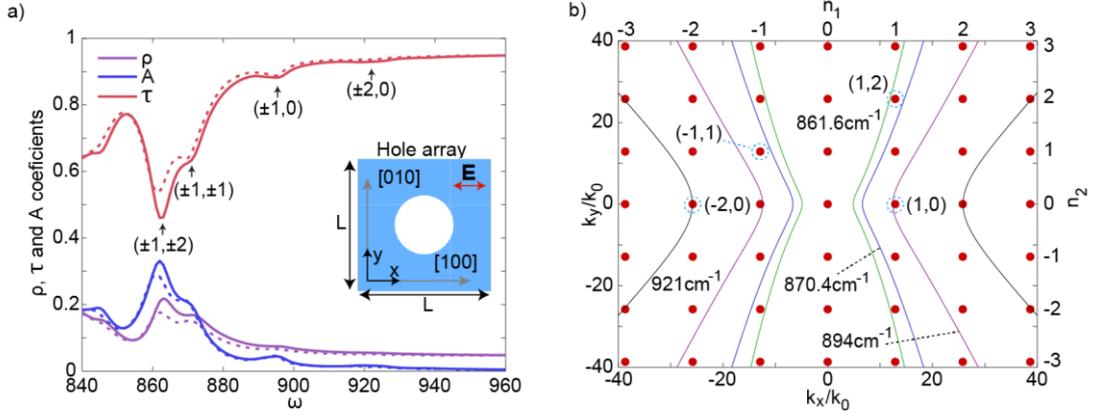

**Figure 5. PhP Bragg resonances in a periodic HA in an α-MoO₃ slab.** (a) Spectra of the transmission, reflection and absorption coefficients ($\tau$, $\rho$ and $A$, respectively). Bragg resonances in different diffraction orders are highlighted by vertical arrows. The inset shows the schematics of a single unit cell. The period of the HA is $L = 900\ nm$ and the hole radius $a = 200\ nm$. The lattice vectors of the HA are aligned with the α-MoO₃ crystallographic axes. The normally-incident light is polarized along the [100] axis. (b) IFCs for several fixed frequencies. The dielectric permittivity of the superstrate is $\varepsilon_1 = 1$, while $\varepsilon_2$ (substrate) is taken for BaF₂.

**References**

1. Basov DN, Fogler MM, Abajo FJGd. Polaritons in van der Waals materials. Science. 2016;354(6309):aag1992.

2. Low T, Chaves A, Caldwell JD, Kumar A, Fang NX, Avouris P, et al. Polaritons in layered two-dimensional materials. Nature Materials. 2017;16(2):182-94.

3. Zhang Q, Hu G, Ma W, Li P, Krasnok A, Hillenbrand R, et al. Interface nano-optics with van der Waals polaritons. Nature. 2021;597(7875):187-95.

4. Dai S, Fei Z, Ma Q, Rodin AS, Wagner M, McLeod AS, et al. Tunable Phonon Polaritons in Atomically Thin van der Waals Crystals of Boron Nitride. Science. 2014;343(6175):1125-9.

5. Ma W, Alonso-González P, Li S, Nikitin AY, Yuan J, Martín-Sánchez J, et al. In-plane anisotropic and ultra-low-loss polaritons in a natural van der Waals crystal.

6. Taboada-Gutiérrez J, Álvarez-Pérez G, Duan J, Ma W, Crowley K, Prieto I, et al. Broad spectral tuning of ultra-low-loss polaritons in a van der Waals crystal by intercalation. Nature Materials. 2020;19(9):964-8.

7. Ma W, Hu G, Hu D, Chen R, Sun T, Zhang X, et al. Ghost hyperbolic surface polaritons in bulk anisotropic crystals. Nature. 2021;596(7872):362-6.

8. Duan J, Capote-Robayna N, Taboada-Gutiérrez J, Álvarez-Pérez G, Prieto I, Martín-Sánchez J, et al. Twisted Nano-Optics: Manipulating Light at the Nanoscale with Twisted Phonon Polaritonic Slabs. Nano Letters. 2020;20(7):5323-9.

9. Duan J, Álvarez-Pérez G, Voronin KV, Prieto I, Taboada-Gutiérrez J, Volkov VS, et al. Enabling propagation of anisotropic polaritons along forbidden directions via a topological transition. Science Advances. 2021;7(14):eabf2690.

10. Duan J, Álvarez-Pérez G, Tresguerres-Mata AIF, Taboada-Gutiérrez J, Voronin KV, Bylinkin A, et al. Planar refraction and lensing of highly confined polaritons in anisotropic media.

11. Alfaro-Mozaz FJ, Rodrigo SG, Vélez S, Dolado I, Govyadinov A, Alonso-González P, et al. Hyperspectral Nanoimaging of van der Waals Polaritonic Crystals. Nano Letters. 2021;21(17):7109-15.

12. Alfaro-Mozaz FJ, Rodrigo SG, Alonso-González P, Vélez S, Dolado I, Casanova F, et al. Deeply subwavelength phonon-polaritonic crystal made of a van der Waals material. Nature Communications. 2019;10(1):42.

13. Yang J, Krix ZE, Kim S, Tang J, Mayyas M, Wang Y, et al. Near-Field Excited Archimedean-like Tiling Patterns in Phonon-Polaritonic Crystals. ACS Nano. 2021;15(5):9134-42.

14. Guddala S, Komissarenko F, Kiriushechkina S, Vakulenko A, Li M, Menon VM, et al. Topological phonon-polariton funneling in midinfrared metasurfaces. Science. 2021;374(6564):225-7.

15. Nikitin AY. Graphene Plasmonics. World Scientific Handbook of Metamaterials and Plasmonics. World Scientific Series in Nanoscience and Nanotechnology, Vol. 4 (Cambridge University Press, Cambridge, 2017), Chap. 8, pp. 307–338.

16. Álvarez-Pérez G, Folland TG, Errea I, Taboada-Gutiérrez J, Duan J, Martín-Sánchez J, et al. Infrared Permittivity of the Biaxial van der Waals Semiconductor α-MoO3 from Near- and Far-Field Correlative Studies. Advanced Materials. 2020;32(29):1908176.

17. Álvarez-Pérez G, Voronin KV, Volkov VS, Alonso-González P, Nikitin AY. Analytical approximations for the dispersion of electromagnetic modes in slabs of biaxial crystals. Physical Review B. 2019;100(23):235408.

18. Sun F, Huang W, Zheng Z, Xu N, Ke Y, Zhan R, et al. Polariton waveguide modes in two-dimensional van der Waals crystals: an analytical model and correlative nano-imaging. Nanoscale. 2021;13(9):4845-54.

19. Slipchenko TM, Nesterov ML, Martin-Moreno L, Nikitin AY. Analytical solution for the diffraction of an electromagnetic wave by a graphene grating. Journal of Optics. 2013;15(11):114008.

20. Hu G, Ou Q, Si G, Wu Y, Wu J, Dai Z, et al. Topological polaritons and photonic magic angles in twisted α-MoO3 bilayers. Nature. 2020;582(7811):209-13.

# Supporting Information:
# Twisted polaritonic crystals in thin van der Waals slabs

Nathaniel Capote-Robayna[1], Olga Matveeva[2], Valentyn S. Volkov[2], Pablo Alonso-González[3,4,] and Alexey Y. Nikitin[1,5,†].

[1]*Donostia International Physics Center (DIPC), Donostia-San Sebastián 20018, Spain.*

[2]*Center for Photonics and 2D Materials, Moscow Institute of Physics and Technology, Dolgoprudny, 141700, Russia.*

[3]*Department of Physics, University of Oviedo, Oviedo 3006, Spain.*

[4]*Center of Research on Nanomaterials and Nanotechnology, CINN (CSIC-Universidad de Oviedo), El entrego 33940, Spain.*

[5]*IKERBASQUE, Basque Foundation for Science, Bilbao 48013,Spain.*

*†Corresponding author. Email: alexey@dipc.org*

## S1. Analytical derivation of the general system of equations for the amplitudes of the scattered fields

In this section we provide a detailed derivation of the linear system of equations 7, 8 presented in the main text. We suppose that the two-dimensional periodic lattice has its basis vectors $\boldsymbol{L}_1$ and $\boldsymbol{L}_2$, as shown on the example of the hole array (HA) in Figure 1a. Each of the lattice vectors can be arbitrarily twisted with respect to the in-plane crystallographic axes. To avoid considering the field inside the slab, we replace the slab by a zero thickness layer parallel to *xy*-plane placed on the substrate, $z = 0$. In this approximation we introduce the in-plane conductivity tensor related to the permittivity of the biaxial slab as follows:

$$\hat{\alpha} = \frac{\pi d}{i\lambda_0}\hat{\varepsilon}, \qquad (S1)$$

where $\hat{\alpha}$ is the effective (2x2) conductivity tensor, $d$ is the slab thickness, $\lambda_0$ is the incident field's wavelength and $\hat{\varepsilon}$ is the permittivity of the material (only the in-plane 2x2 in-plane sub-tensor of the whole 3x3 $\hat{\varepsilon}$ tensor appears in this equation). The assumed periodicity of the slab implies that the effective conductivity tensor $\hat{\alpha}$ is also periodic along the translation vectors $\boldsymbol{L}_1 = (L_{1x}, L_{1y}, 0)$ and $\boldsymbol{L}_2 = (L_{2x}, L_{2y}, 0)$, which mathematically can be written as:

$$\hat{\alpha}(\boldsymbol{r}) = \hat{\alpha}(\boldsymbol{r} + n_1 \boldsymbol{L}_1 + n_2 \boldsymbol{L}_2), \tag{S2}$$

The Fourier expansion (FE) of $\hat{\alpha}(\boldsymbol{r})$ can be expressed as follows:

$$\hat{\alpha}(\boldsymbol{r}) = \sum_{n_1 n_2} \tilde{\alpha}_{n_1 n_2} e^{i(n_1 \boldsymbol{g}_1 + n_2 \boldsymbol{g}_2)\boldsymbol{r}}, \tag{S3}$$

being $\tilde{\alpha}_{n_1 n_2}$ the Fourier expansion amplitude and $\boldsymbol{g}_1 = (g_{1x}, g_{1y}, 0)$ and $\boldsymbol{g}_2 = (g_{2x}, g_{2y}, 0)$ the reciprocal space vectors, the latter satisfying $\boldsymbol{g}_i \boldsymbol{L}_j = 2\pi \delta_{ij}$. To simplify the notations, we will a use multi-index, $N$, containing both $n_1$ and $n_2$ ($N \equiv \{n_1, n_2\}$). The matrix form of the Fourier amplitudes reads:

$$\tilde{\alpha}_N = \begin{pmatrix} \alpha_N^{xx} & \alpha_N^{xy} \\ \alpha_N^{xy} & \alpha_N^{yy} \end{pmatrix}. \tag{S4}$$

Note that if $\hat{\alpha}(\boldsymbol{r})$ is symmetric, then $\tilde{\alpha}_N$ is symmetric as well.

To describe the electromagnetic fields in each media we will use the following basis vectors (being, in general, quasi-orthogonal modes in case of a weakly-absorbing medium) for s- and p- polarized light (in Dirac notations):

$$|s1N\rangle = \frac{1}{k_{tN}} \begin{pmatrix} -k_{yN} \\ k_{xN} \\ 0 \end{pmatrix} e^{i\boldsymbol{r}\boldsymbol{k}_{tN}}, \tag{S5}$$

$$|p1N\rangle = \frac{1}{k_{tN}} \begin{pmatrix} k_{xN} \\ k_{yN} \\ k_{tN}^2 \\ -k_{z1N} \end{pmatrix} e^{i\boldsymbol{r}\boldsymbol{k}_{tN}}, \tag{S6}$$

$$|s2N\rangle = \frac{1}{k_{tN}} \begin{pmatrix} -k_{yN} \\ k_{xN} \\ 0 \end{pmatrix} e^{i\boldsymbol{r}\boldsymbol{k}_{tN}}, \tag{S7}$$

$$|p2N\rangle = \frac{1}{k_{tN}} \begin{pmatrix} k_{xN} \\ k_{yN} \\ k_{tN}^2 \\ k_{z2N} \end{pmatrix} e^{i\boldsymbol{r}\boldsymbol{k}_{tN}}, \quad (S8)$$

where the index $i = \{1,2\}$ refers to the media (dielectric half-space) and the wavevectors of the electromagnetic modes are given by:

$$\boldsymbol{k}_t = (k_x, k_y, 0), \quad (S9)$$

$$k_{xN} = k_x + n_1 g_{1x} + n_2 g_{2x}, \quad (S10)$$

$$k_{yN} = k_y + n_1 g_{1y} + n_2 g_{2y}, \quad (S11)$$

$$\boldsymbol{k}_{tN} = \boldsymbol{k}_t + n_1 \boldsymbol{g}_1 + n_2 \boldsymbol{g}_2 = (k_{xN}, k_{yN}, 0), \quad (S12)$$

$$k_{z1N} = \sqrt{\varepsilon_1 k_0^2 - \boldsymbol{k}_{tN}^2}, \quad (S13)$$

$$k_{z2N} = \sqrt{\varepsilon_2 k_0^2 - \boldsymbol{k}_{tN}^2}. \quad (S14)$$

To match the fields by the boundary conditions, we will use the in-plane projections of the basis vectors:

$$|sN\rangle = \frac{1}{k_{tN}} \begin{pmatrix} -k_{yN} \\ k_{xN} \end{pmatrix} e^{i\boldsymbol{r}\boldsymbol{k}_{tN}}, \quad (S15)$$

$$|pN\rangle = \frac{1}{k_{tN}} \begin{pmatrix} k_{xN} \\ k_{yN} \end{pmatrix} e^{i\boldsymbol{r}\boldsymbol{k}_{tN}}. \quad (S16)$$

The scalar products of the basis vectors, $\langle \beta'N'|\beta N\rangle$, possess orthogonality. In order to make sure in the latter property, we can choose the lattice vectors as $\boldsymbol{L}_1 = (L_x, 0, 0)$ and $\boldsymbol{L}_2 = (L_{2x}, L_y, 0)$. The reciprocal space vectors in this case read as follows:

$$\boldsymbol{g}_1 = \left(\frac{2\pi}{L_x}, -\frac{2\pi L_{2x}}{L_y L_x}, 0\right), \quad \boldsymbol{g}_2 = \left(0, \frac{2\pi}{L_y}, 0\right). \quad (S17)$$

Thus, the scalar product can be explicitly written:

$$\langle \beta'N'|\beta N\rangle = \frac{1}{L_xL_y}\left(\beta'_{xN'}, \beta'_{yN'}\right)\begin{pmatrix}\beta_{xN}\\\beta_{yN}\end{pmatrix}\int_{-\frac{L_y}{2}}^{\frac{L_y}{2}}dy\int_{-\frac{L_x}{2}}^{\frac{L_x}{2}}dx\, e^{i\left((n_1-n'_1)g_1+(n_2-n'_2)g_2\right)r} =$$

$$\frac{1}{L_xL_y}\left(\beta'_{xN'}, \beta'_{yN'}\right)\begin{pmatrix}\beta_{xN}\\\beta_{yN}\end{pmatrix}\int_{-\frac{L_y}{2}}^{\frac{L_y}{2}}dy\int_{-\frac{L_x}{2}}^{\frac{L_x}{2}}dx\, e^{i\left[\frac{2\pi}{L_x}(n_1-n'_1)x+\left(\frac{2\pi}{L_y}(n_2-n'_2)-\frac{2\pi L_{2x}}{L_yL_x}(n_1-n'_1)\right)y\right]} =$$

$$\frac{1}{L_y}\left(\beta'_{xN'}, \beta'_{yN'}\right)\begin{pmatrix}\beta_{xN}\\\beta_{yN}\end{pmatrix}\frac{e^{i\pi(n_1-n'_1)}-e^{i\pi(n_1-n'_1)}}{2\pi i(n_1-n'_1)}\int_{-\frac{L_y}{2}}^{\frac{L_y}{2}}dy\, e^{i\left(\frac{2\pi}{L_y}(n_2-n'_2)-\frac{2\pi L_{2x}}{L_yL_x}(n_1-n'_1)\right)y} =$$

$$\frac{1}{L_y}\left(\beta'_{xN'}, \beta'_{yN'}\right)\begin{pmatrix}\beta_{xN}\\\beta_{yN}\end{pmatrix}\delta_{n_1n'_1}\int_{-\frac{L_y}{2}}^{\frac{L_y}{2}}dy\, e^{i\left(\frac{2\pi}{L_y}(n_2-n'_2)-\frac{2\pi L_{2x}}{L_yL_x}(n_1-n'_1)\right)y} =$$

$$\frac{1}{L_y}\left(\beta'_{xN'}, \beta'_{yN'}\right)\begin{pmatrix}\beta_{xN}\\\beta_{yN}\end{pmatrix}\delta_{n_1n'_1}\int_{-\frac{L_y}{2}}^{\frac{L_y}{2}}dy\, e^{i\left(\frac{2\pi}{L_y}(n_2-n'_2)\right)y} =$$

$$\left(\beta'_{xN'}, \beta'_{yN'}\right)\begin{pmatrix}\beta_{xN}\\\beta_{yN}\end{pmatrix}\delta_{n_1n'_1}\frac{e^{i\pi(n_2-n'_2)}-e^{i\pi(n_2-n'_2)}}{2\pi i(n_2-n'_2)} =$$

$$\left(\beta'_{xN'}, \beta'_{yN'}\right)\begin{pmatrix}\beta_{xN}\\\beta_{yN}\end{pmatrix}\delta_{n_1n'_1}\delta_{n_2n'_2}. \tag{S18}$$

So that, we have proved that:

$$\langle \beta'N'|\beta N\rangle = \delta_{\beta\beta'}\delta_{NN'}. \tag{S19}$$

Using the above basis vectors, the electric field in each medium can be represented as a superposition of plane waves:

$$\boldsymbol{E}_1(z) = \left(I_s|s10\rangle + I_p|p10\rangle\right)e^{-izk_{inc}} +$$
$$\sum_N\left[(R_{sN}|s1N\rangle + R_{pN}|p1N\rangle)e^{-izk_{z1N}}\right] \qquad z\geq 0, \tag{S20}$$

$$\boldsymbol{E}_2(z) = \sum_N\left[(T_{sN}|s2N\rangle + T_{pN}|p2N\rangle)e^{izk_{z2N}}\right] \qquad z<0, \tag{S21}$$

where I, R and T stand for amplitudes of the modes for the incident, "reflected" and "transmitted" electric fields and indices s and p indicate the polarization. The out-of-plane momentum of the incident wave is $k_{inc} = \sqrt{\varepsilon_1 k_0^2 - \boldsymbol{k}_{t00}^2}$.

The boundary conditions for the electromagnetic fields at $z = 0$ read:

$$\boldsymbol{E}_{1t}(z = 0) = \boldsymbol{E}_{2t}(z = 0), \tag{S22}$$

$$\boldsymbol{e}_z \times (\boldsymbol{H}_{1t}(z = 0) - \boldsymbol{H}_{2t}(z = 0)) = 2\hat{\alpha}\boldsymbol{E}_{1t}(z = 0), \tag{S23}$$

where the subscript t, means the in-plane (transversal) components, while $\boldsymbol{e}_z$ is the unitary vector along the z-direction. Substituting the electric fields from Eqs S20, S21 into Eq S22 we obtain:

$$I_s|s00\rangle + I_p|p00\rangle + \sum_N (R_{sN}|sN\rangle + R_{pN}|pN\rangle) = \sum_N (T_{sN}|sN\rangle + T_{pN}|pN\rangle). \tag{S24}$$

Grouping and projecting by bras $\langle sN|$ and $\langle pN|$ we get:

$$I_s\langle sN|s00\rangle + I_p\langle sN|p00\rangle +$$
$$\sum_N [(R_{sN} - T_{sN})\langle sN|sN\rangle + (R_{pN} - T_{pN})\langle sN|pN\rangle] = 0, \tag{S25}$$

$$I_s\langle pN|s00\rangle + I_p\langle pN|p00\rangle +$$
$$\sum_N [(R_{sN} - T_{sN})\langle pN|sN\rangle + (R_{pN} - T_{pN})\langle pN|pN\rangle] = 0. \tag{S26}$$

Using the orthogonality of the basis vectors from Eq S19, equations S25, S26 simplify to:

$$I_\beta \delta_{N00} + R_{\beta N} = T_{\beta N}, \tag{S27}$$

for all possible multi-index values, $N$.

Now we proceed with the boundary condition for magnetic fields. The magnetic field can be described as a function of the electric field as $\boldsymbol{H} = \boldsymbol{q} \times \boldsymbol{E}$, where $\boldsymbol{q} = \boldsymbol{k}/k_0$. So that, applying the vector product, the following identities are obtained:

$$\mathbf{e}_z \times \mathbf{q}_{iN} \times |\beta iN\rangle_t = \mp Y_{iN}^{\beta}|\beta N\rangle, \tag{S28}$$

where $Y_{iN}^s = q_{ziN}$ and $Y_{iN}^p = \frac{\varepsilon_i}{q_{ziN}}$, while $Y_{inc}^{\beta'}$ takes value $Y_{inc}^s = k_{inc}/k_0$ and $Y_{inc}^p = \varepsilon_1 k_0/k_{inc}$. Substituting expression S28 into S23 we obtain:

$$I_s Y_{inc}^s |s00\rangle + I_p Y_{inc}^p |p00\rangle - \sum_N \left(R_{sN} Y_{1N}^s |sN\rangle + R_{pN} Y_{1N}^p |pN\rangle\right) -$$

$$\sum_N \left(T_{sN} Y_{2N}^s |sN\rangle + T_{pN} Y_{2N}^p |pN\rangle\right) =$$

$$2\hat{\alpha} \sum_N \left(T_{sN} |sN\rangle + T_{pN} |pN\rangle\right). \tag{S29}$$

Grouping all terms and projecting by bras $\langle sN'|$ and $\langle pN'|$, we arrive at:

$$\sum_N \left[(R_{sN} Y_{1N}^s + T_{sN} Y_{2N}^s)\delta_{NN'} + 2T_{sN}\langle sN'|\hat{\alpha}|sN\rangle + 2T_{pN}\langle sN'|\hat{\alpha}|pN\rangle\right] =$$

$$I_s Y_{inc}^s \delta_{00N'}, \tag{S30}$$

$$\sum_N \left[2T_{sN}\langle pN'|\hat{\alpha}|sN\rangle + (R_{pN} Y_{1N}^p + T_{pN} Y_{2N}^p)\delta_{NN'} + 2T_{pN}\langle pN'|\hat{\alpha}|pN\rangle\right] =$$

$$I_p Y_{inc}^p \delta_{00N'}. \tag{S31}$$

Finally, using the previously derived identity S27 and introducing the matrix

$$M_{NN'}^{\beta\beta'} \equiv \langle \beta'N'|\hat{\alpha}|\beta N\rangle, \tag{S32}$$

the system of equations can be written compactly:

$$\sum_{\beta N} D_{NN'}^{\beta\beta'} T_{\beta N} = 2 I_{\beta'} Y_{inc}^{\beta'} \delta_{00N'}. \tag{S33}$$

$$D_{NN'}^{\beta\beta'} = \left(Y_{1N}^{\beta} + Y_{2N}^{\beta}\right)\delta_{\beta\beta'}\delta_{NN'} + 2M_{NN'}^{\beta\beta'}. \tag{S34}$$

The matrix elements of $M_{NN'}^{\beta\beta'}$ read explicitly:

$$\langle \beta' N' | \hat{\alpha} | \beta N \rangle =$$

$$\frac{1}{L_x L_y} \int_{-\frac{L_y}{2}}^{\frac{L_y}{2}} dy \int_{-\frac{L_x}{2}}^{\frac{L_x}{2}} dx \left[ \left( \beta'_{xN'}, \beta'_{yN'} \right) \hat{\alpha} \begin{pmatrix} \beta_{xN} \\ \beta_{yN} \end{pmatrix} e^{i((n_1-n'_1)g_1+(n_2-n'_2)g_2)r} \right] =$$

$$\frac{1}{L_x L_y} \int_{-\frac{L_y}{2}}^{\frac{L_y}{2}} dy \int_{-\frac{L_x}{2}}^{\frac{L_x}{2}} dx \sum_{N''} \left[ \left( \beta'_{xN'}, \beta'_{yN'} \right) \tilde{\alpha}_{N''} \begin{pmatrix} \beta_{xN} \\ \beta_{yN} \end{pmatrix} e^{i((n''_1+n_1-n'_1)g_1+(n''_2+n_2-n'_2)g_2)r} \right] =$$

$$\sum_{N''} \left[ \left( \beta'_{xN'}, \beta'_{yN'} \right) \tilde{\alpha}_{N''} \begin{pmatrix} \beta_{xN} \\ \beta_{yN} \end{pmatrix} \frac{1}{L_x L_y} \int_{-\frac{L_y}{2}}^{\frac{L_y}{2}} dy \int_{-\frac{L_x}{2}}^{\frac{L_x}{2}} dx \, e^{i((n''_1+n_1-n'_1)g_1+(n''_2+n_2-n'_2)g_2)r} \right]. \quad (S35)$$

The integrals in S35 can be taken using the same vector basis as in Eq S18, leading to the following expression:

$$\frac{1}{L_x L_y} \int_{-\frac{L_y}{2}}^{\frac{L_y}{2}} dy \int_{-\frac{L_x}{2}}^{\frac{L_x}{2}} dx \, e^{i((n''_1+n_1-n'_1)g_1+(n''_2+n_2-n'_2)g_2)r} =$$

$$\frac{1}{L_x L_y} \int_{-\frac{L_y}{2}}^{\frac{L_y}{2}} dy \int_{-\frac{L_x}{2}}^{\frac{L_x}{2}} dx \, e^{i\left[\frac{2\pi}{L_x}(n''_1+n_1-n'_1)x + \left(\frac{2\pi}{L_y}(n''_2+n_2-n'_2) - \frac{2\pi L_{2x}}{L_y L_x}(n''_1+n_1-n'_1)\right)y\right]} =$$

$$\frac{1}{L_y} \frac{e^{i\pi(n''_1+n_1-n'_1)} - e^{i\pi(n''_1+n_1-n'_1)}}{2\pi i (n''_1+n_1-n'_1)} \int_{-\frac{L_y}{2}}^{\frac{L_y}{2}} dy \, e^{i\left(\frac{2\pi}{L_y}(n''_2+n_2-n'_2) - \frac{2\pi L_{2x}}{L_y L_x}(n''_1+n_1-n'_1)\right)y} =$$

$$\frac{1}{L_y} \delta_{n''_1 n'_1 - n_1} \int_{-\frac{L_y}{2}}^{\frac{L_y}{2}} dy \, e^{i\left(\frac{2\pi}{L_y}(n''_2+n_2-n'_2) - \frac{2\pi L_{2x}}{L_y L_x}(n''_1+n_1-n'_1)\right)y} =$$

$$\frac{1}{L_y} \delta_{n''_1 n'_1 - n_1} \int_{-\frac{L_y}{2}}^{\frac{L_y}{2}} dy \, e^{i\left(\frac{2\pi}{L_y}(n''_2+n_2-n'_2)\right)y} =$$

$$\delta_{n''_1 n'_1 - n_1} \frac{e^{i\pi(n''_2+n_2-n'_2)} - e^{i\pi(n''_2+n_2-n'_2)}}{2\pi i (n''_2+n_2-n'_2)} = \delta_{n''_1 n'_1 - n_1} \delta_{n''_2 n'_2 - n_2}. \quad (S36)$$

Thus, substituting the result of the integration from Eq S36 into S35, we get:

$$\langle \beta'N'|\alpha|\beta N\rangle =$$

$$\sum_{N''}\left[\left(\beta'_{xN'},\beta'_{yN'}\right)\tilde{\alpha}_{N''}\begin{pmatrix}\beta_{xN}\\\beta_{yN}\end{pmatrix}\frac{1}{L_xL_y}\int_{-\frac{L_y}{2}}^{\frac{L_y}{2}}dy\int_{-\frac{L_x}{2}}^{\frac{L_x}{2}}dx\, e^{i((n_1''+n_1-n_1')g_1+(n_2''+n_2-n_2')g_2)r}\right]=$$

$$\sum_{N''}\left[\left(\beta'_{xN'},\beta'_{yN'}\right)\tilde{\alpha}_{N''}\begin{pmatrix}\beta_{xN}\\\beta_{yN}\end{pmatrix}\delta_{n_1''n_1'-n_1}\delta_{n_2''n_2'-n_2}\right]=$$

$$\left(\beta'_{xN'},\beta'_{yN'}\right)\tilde{\alpha}_{N'-N}\begin{pmatrix}\beta_{xN}\\\beta_{yN}\end{pmatrix}. \tag{S37}$$

Then the matrix elements $M_{NN'}^{\beta\beta'}$ read explicitly as:

$$M_{NN'}^{ss}=\frac{1}{k_{tN}k_{tN'}}(-k_{yN'},k_{xN'})\begin{pmatrix}\tilde{\alpha}_{N'-N}^{xx}&\tilde{\alpha}_{N'-N}^{xy}\\\tilde{\alpha}_{N'-N}^{xy}&\tilde{\alpha}_{N'-N}^{yy}\end{pmatrix}\begin{pmatrix}-k_{yN}\\k_{xN}\end{pmatrix}=$$

$$\frac{\tilde{\alpha}_{N'-N}^{xx}k_{yN}k_{yN'}+\tilde{\alpha}_{N'-N}^{xy}(-k_{yN}k_{xN'}-k_{xN}k_{yN'})+\tilde{\alpha}_{N'-N}^{yy}k_{xN}k_{xN'}}{k_{tN}k_{tN'}}, \tag{S38}$$

$$M_{NN'}^{sp}=\frac{1}{k_{tN}k_{tN'}}(-k_{yN'},k_{xN'})\begin{pmatrix}\tilde{\alpha}_{N'-N}^{xx}&\tilde{\alpha}_{N'-N}^{xy}\\\tilde{\alpha}_{N'-N}^{xy}&\tilde{\alpha}_{N'-N}^{yy}\end{pmatrix}\begin{pmatrix}k_{xN}\\k_{yN}\end{pmatrix}=$$

$$\frac{-\tilde{\alpha}_{N'-N}^{xx}k_{xN}k_{yN'}+\tilde{\alpha}_{N'-N}^{xy}(k_{xN}k_{xN'}-k_{yN}k_{yN'})+\tilde{\alpha}_{N'-N}^{yy}k_{yN}k_{xN'}}{k_{tN}k_{tN'}}, \tag{S39}$$

$$M_{NN'}^{ps}=\frac{1}{k_{tN}k_{tN'}}(k_{xN'},k_{yN'})\begin{pmatrix}\tilde{\alpha}_{N'-N}^{xx}&\tilde{\alpha}_{N'-N}^{xy}\\\tilde{\alpha}_{N'-N}^{xy}&\tilde{\alpha}_{N'-N}^{yy}\end{pmatrix}\begin{pmatrix}-k_{yN}\\k_{xN}\end{pmatrix}=$$

$$\frac{-\tilde{\alpha}_{N'-N}^{xx}k_{yN}k_{xN'}+\tilde{\alpha}_{N'-N}^{xy}(k_{xN}k_{xN'}-k_{yN}k_{yN'})+\tilde{\alpha}_{N'-N}^{yy}k_{xN}k_{yN'}}{k_{tN}k_{tN'}}, \tag{S40}$$

$$M_{NN'}^{pp}=\frac{1}{k_{tN}k_{tN'}}(k_{xN'},k_{yN'})\begin{pmatrix}\tilde{\alpha}_{N'-N}^{xx}&\tilde{\alpha}_{N'-N}^{xy}\\\tilde{\alpha}_{N'-N}^{xy}&\tilde{\alpha}_{N'-N}^{yy}\end{pmatrix}\begin{pmatrix}k_{xN}\\k_{yN}\end{pmatrix}=$$

$$\frac{\tilde{\alpha}_{N'-N}^{xx}k_{xN}k_{xN'}+\tilde{\alpha}_{N'-N}^{xy}(k_{yN}k_{xN'}-k_{xN}k_{yN'})+\tilde{\alpha}_{N'-N}^{yy}k_{yN}k_{yN'}}{k_{tN}k_{tN'}}. \tag{S41}$$

Furthermore, the matrix elements $M_{NN'}^{\beta\beta'}$ given by the expressions S38-S41 can be compacted as follows:

$$M_{NN'}^{\beta\beta'} =$$

$$\frac{1}{k_{tN}k_{tN'}}\Big[(-1)^{1+\delta_{\beta\beta'}}\alpha_{N'-N}^{xx}k_{\bar{\gamma}N}k_{\bar{\gamma}'N'} +$$

$$\alpha_{N'-N}^{xy}\Big((-1)^{1+\delta_{\beta p}}k_{\gamma'N}k_{\bar{\gamma}N'} + (-1)^{1+\delta_{p\beta'}}k_{\bar{\gamma}'N}k_{\gamma N'}\Big) +$$

$$\alpha_{N'-N}^{yy}k_{\gamma N}k_{\gamma'N'}\Big], \qquad (S42)$$

where $\gamma$ stays for $x$ or $y$ when $\beta$ takes s and p values respectively, while oppositely $\bar{\gamma}$ stays for $x$ or $y$ when $\beta$ takes p and s values, respectively. Note that $\gamma$ is a function of $\beta$, while $\gamma'$ depends upon $\beta'$.

## S2. Simplified solution for the harmonic lattices

Here we further simplify the system of equation and its solution for a particular case of the first-order resonance at normal incidence. Namely, we assume that the resonance condition is simultaneously fulfilled for four lowest diffraction orders: (-1,0), (1,0), (0,-1), (0,1). As the main simplification, we consider only these resonant diffraction orders together with (0,0) (neglecting the rest of the diffraction orders), thus restricting the diffraction on the lattice by the interaction between the five lowest diffracted orders. Furthermore, we neglect contributions from the s-polarization components of the electric field. This neglection can be justified by the fact that the emergent resonance is due to the excitation of p-polarized plane waves (polaritons). As a result, in the whole system of equations we retain only the amplitudes $T_{p10}$, $T_{p01}$, $T_{p-10}$, $T_{p0-1}$ and $T_{p00}$. For compactness, the notation for the matrix $M_{n_1'n_1n_2'n_2}^{pp}$ will be simplified to $M_{n_2'n_2}^{n_1'n_1}$ (since the index of the polarization does not introduce any additional information). Let us explicitly write the five equations for the five unknown amplitudes from the system S33:

- $n_1 = 0$ and $n_2 = \pm 1$:
$$2M_{0\pm1}^{00}T_{p00} + \big(Y_{10\pm1}^{p} + Y_{20\pm1}^{p} + 2M_{\pm1\pm1}^{00}\big)T_{p0\pm1} = 0, \qquad (S43)$$

- $n_1 = \pm 1$ and $n_2 = 0$:
$$2M_{00}^{0\pm1}T_{p00} + \big(Y_{1\pm10}^{p} + Y_{2\pm10}^{p} + 2M_{00}^{\pm1\pm1}\big)T_{p\pm10} = 0, \qquad (S44)$$

- $n_1 = 0$ and $n_2 = 0$:
$$2M^{00}_{-10}T_{p0-1} + 2M^{-10}_{00}T_{p-10} + \left(Y^p_{100} + Y^p_{200} + 2M^{00}_{00}\right)T_{p00} + 2M^{00}_{10}T_{p01} +$$
$$2M^{10}_{00}T_{p10} = 2Y^p_{100}. \tag{S45}$$

Using the symmetry property, $M^{n'_1 n_1}_{n'_2 n_2} = M^{n_1 n'_1}_{n_2 n'_2}$, and introducing the following notation:
$$b_{n_1 n_2} = Y^p_{1 n_1 n_2} + Y^p_{2 n_1 n_2} + 2M^{n_1 n_1}_{n_2 n_2}, \tag{S46}$$
the equations S43-S45 can be rewritten as:

- $n_1 = 0$ and $n_2 = \pm 1$:
$$2M^{00}_{\pm 10}T_{p00} + b_{0 \pm 1}T_{p01} = 0, \tag{S47}$$

- $n_1 = \pm 1$ and $n_2 = 0$:
$$2M^{\pm 10}_{00}T_{p00} + b_{\pm 10}T_{p \pm 10} = 0, \tag{S48}$$

- $n_1 = 0$ and $n_2 = 0$:
$$2M^{00}_{-10}T_{p0-1} + 2M^{-10}_{00}T_{p-10} + b_{00}T_{p00} + 2M^{00}_{10}T_{p01} + 2M^{10}_{00}T_{p10} =$$
$$2Y^p_{100}. \tag{S49}$$

The amplitudes $T_{p0 \pm 1}$ and $T_{p \pm 10}$ can be written as a function of $T_{p00}$:

$$T_{p0 \pm 1} = -\frac{2M^{00}_{\pm 10}}{b_{0 \pm 1}}T_{p00}, \tag{S50}$$

$$T_{p \pm 10} = -\frac{2M^{\pm 10}_{00}}{b_{\pm 10}}T_{p00}. \tag{S51}$$

Substituting Eqs S50, S51 into S46 we obtain the equation for $T_{p00}$:

$$-\frac{4(M^{00}_{-10})^2}{b_{0-1}}T_{p00} - \frac{4(M^{-10}_{00})^2}{b_{-10}}T_{p00} + b_{00}T_{p00} - \frac{4(M^{00}_{10})^2}{b_{01}}T_{p00} -$$
$$\frac{4(M^{10}_{00})^2}{b_{10}}T_{p00} = 2Y^p_{100}. \tag{S52}$$

We find explicitly from S52:

$$T_{p00} = \frac{2Y^p_{100}}{\Lambda b_{00}}, \tag{S53}$$

where:

$$\Lambda = 1 - \frac{4}{b_{00}}\left(\frac{(M^{00}_{10})^2}{b_{0-1}} + \frac{(M^{-10}_{00})^2}{b_{-10}} + \frac{(M^{00}_{10})^2}{b_{01}} + \frac{(M^{10}_{00})^2}{b_{10}}\right). \tag{S54}$$

Now the rest of the amplitudes can be found from Eqs S50, S51 using S53:

$$T_{p0\pm1} = -\frac{4Y_{100}^{p} M_{\pm10}^{00}}{\Lambda b_{00} b_{0\pm1}},\tag{S55}$$

$$T_{p\pm10} = -\frac{4Y_{100}^{p} M_{00}^{\pm10}}{\Lambda b_{00} b_{\pm10}}.\tag{S56}$$

For further simplifications, we will use the assumption that the period of the lattice is much smaller than the wavelength of the incident light, $L \ll \lambda$. Then $k_x, k_y \ll g_1, g_2$, for all orders different from $n_1 = n_2 = 0$. So that, first order wavevectors read as:

$$\boldsymbol{k}_{tN} \approx n_1 \boldsymbol{g}_1 + n_2 \boldsymbol{g}_2,\tag{S57}$$

$$k_{zjN} \approx k_0 \sqrt{\varepsilon_j - (n_1^2 + n_2^2)\lambda^2/L^2} \approx ik_0 \lambda/L \sqrt{n_1^2 + n_2^2},\tag{S58}$$

Additionally, recall that we assume the normally-incident illumination, and thus

$$k_{zj00} = k_0 \sqrt{\varepsilon_j}.\tag{S59}$$

With these simplifications, the coefficients $Y_{jn_1n_2}^{p}$ become:

$$Y_{j00}^{p} = \sqrt{\varepsilon_j},\tag{S60}$$

$$Y_{j\pm10}^{p} = Y_{j0\pm1}^{p} \approx -i\frac{\varepsilon_j L}{\lambda},\tag{S61}$$

To further simplify the matrix elements, we also assume that the incident wave is polarized along the $x$-axis. Considering that $k_{x00} = 0$, the matrix elements read:

$$M_{00}^{00} \approx \alpha_{00}^{xx},\tag{S62}$$

$$M_{00}^{0\pm1} = M_{00}^{\pm10} \approx \alpha_{\pm10}^{xx},\tag{S63}$$

$$M_{00}^{\pm1\pm1} \approx \alpha_{00}^{xx},\tag{S64}$$

$$M_{0\pm1}^{00} \approx -\alpha_{0\pm1}^{xy},\tag{S65}$$

$$M_{\pm10}^{00} \approx \alpha_{0\pm1}^{xy},\tag{S66}$$

$$M_{\pm1\pm1}^{00} \approx \alpha_{00}^{yy},\tag{S67}$$

Substituting S60-S67 into $b_{n_1n_2}$, we have:

$$b_{00} \approx 2\alpha_{00}^{xx} + \sqrt{\varepsilon_1} + \sqrt{\varepsilon_2}, \qquad (S68)$$

$$b_{\pm 10} \approx 2\alpha_{00}^{xx} - i\frac{L}{\lambda}(\varepsilon_1 + \varepsilon_2), \qquad (S69)$$

$$b_{0\pm 1} \approx 2\alpha_{00}^{yy} - i\frac{L}{\lambda}(\varepsilon_1 + \varepsilon_2), \qquad (S70)$$

The term $\Lambda$ simplifies as:

$$\Lambda \approx 1 - \frac{4}{2\alpha_{00}^{xx} + \sqrt{\varepsilon_1} + \sqrt{\varepsilon_2}}\left(\frac{(\alpha_{0-1}^{xy})^2}{b_{0-1}} + \frac{(\alpha_{-10}^{xx})^2}{b_{-10}} + \frac{(\alpha_{10}^{xx})^2}{b_{10}} + \frac{(\alpha_{01}^{xy})^2}{b_{01}}\right). \qquad (S71)$$

Substituting S60-S71 into the expressions S53, S55, S56, we obtain the simplified expressions for the resonant and zero-order amplitudes:

$$T_{p00} = \frac{2\sqrt{\varepsilon_1}}{\Lambda[\sqrt{\varepsilon_1} + \sqrt{\varepsilon_2} + 2\alpha_{00}^{xx}]}. \qquad (S72)$$

$$T_{p0\pm 1} = \frac{4\sqrt{\varepsilon_1}\alpha_{01}^{xy}}{\Lambda[\sqrt{\varepsilon_1} + \sqrt{\varepsilon_2} + 2\alpha_{00}^{xx}]\left[2\alpha_{00}^{yy} - i\frac{L}{\lambda}(\varepsilon_1 + \varepsilon_2)\right]}, \qquad (S73)$$

$$T_{p\pm 10} = \frac{4\sqrt{\varepsilon_1}\alpha_{10}^{xx}}{\Lambda[\sqrt{\varepsilon_1} + \sqrt{\varepsilon_2} + 2\alpha_{00}^{xx}]\left[2\alpha_{00}^{xx} - i\frac{L}{\lambda}(\varepsilon_1 + \varepsilon_2)\right]}. \qquad (S74)$$

Note that using a similar analytical procedure, it is possible to obtain explicit expressions for the diffraction amplitudes in case of the resonance in other (higher) orders, as well as for arbitrary twisting angles.

### S3. Convergence of FT harmonics and transmittance coefficient

The linear system of equations S33 is infinite, and therefore for the numerical solution it has to be truncated up to an order $n_{max}$ (where $n_{max}$ means that $n_1, n_2 \in \{-n_{max}, \ldots, 0, \ldots, n_{max}\}$). This $n_{max}$ order should be, on the one hand, large enough to guarantee convergence of the solution, but on the other hand small enough to provide a reasonable calculation time. In this this section we analyze the convergence of the FE for conductivity matrix tensor and transmission coefficient.

The spatial dependence of the conductivity tensor can be expressed as:

$$\hat{\alpha}(\mathbf{r}) = \hat{\alpha}(1 - f(\mathbf{r})), \qquad (S75)$$

where $f$ takes 1 (0) value inside (outside) of the holes with radius $a$, respectively. The FE of $f(\mathbf{r})$ in case of a rectangular array (with period $L$) of circular holes is:

$$f_{n_1 n_2} = \frac{a}{L} \frac{1}{\sqrt{n_1^2 + n_2^2}} J_1\left(2\pi \frac{a}{L}\sqrt{n_1^2 + n_2^2}\right), \qquad (S76)$$

being $J_1$ the Bessel function of the first kind. Substituting S76 into S75 we obtain an expression for the FE coefficients, $\hat{\alpha}_{n_1,n_2}$:

$$\hat{\alpha}_{n_1 n_2} = \hat{\alpha}\left[\delta_{n_1 0}\delta_{n_2 0} - \frac{a}{L}\frac{1}{\sqrt{n_1^2 + n_2^2}} J_1\left(2\pi \frac{a}{L}\sqrt{n_1^2 + n_2^2}\right)\right]. \qquad (S77)$$

For simplicity, we plot in Figure S1a the function $f(\mathbf{r})$ "reconstructed" from its FE, for different truncations of the Fourier harmonics, (from $n_{max} = 1$ to $n_{max} = 35$), together with the original (exact) $f(\mathbf{r})$. As it follows from the colorplots, the larger the truncation number, $n_{max}$ (and consequently, the amount of the retained harmonics), the closer is the reconstructed function to its original. For $n_{max} = 35$ the reconstructed image is virtually indistinguishable from the original function. To make sure that the latter truncation indeed guarantees the convergence of the numeric solution of the system of equations S33, in Figure S1b we represent the transmission coefficient, $\tau = \sum_{\beta N} Re(Y_{2N}^{\beta}/Y_{inc}^{\beta})|T_{\beta N}|^2$, as a function of the $n_{max}$, for several frequencies from the spectral range considered in Figure 5 of the main text. In the system of equations S33 the number of the field Fourier harmonics is truncated consistently, up to $n_{max} = 35$. As it follows from Figure S1b, the convergence of the solution is achieved for $n_{max} = 75$, although already for $n_{max} = 35$ the deviation from the convergent value is of order of 2%. Therefore, the latter truncation was taken for the numeric simulations shown in the main text.

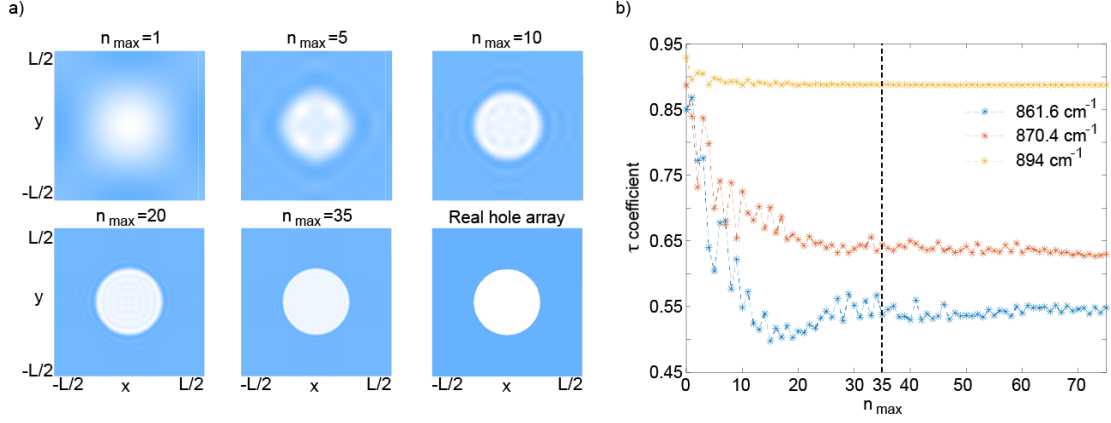

**Figure S1. Convergence of the numeric solution as function of the number of the retained Fourier harmonics. (a)** Schematics of the Fourier expansion of the function $f(r)$ inside the Wigner-Seitz cell (white for minimum value and blue for maximum value). The reconstructed function $f_{n_1 n_2}$ is shown for $n_{max} = 1, 5, 10, 20$, and 35. These cases are compared with the step function $f(r)$, in the last panel. **(b)** Transmission coefficient, $\tau$, as a function of $n_{max}$ for the square HA with the period $L = 900\ nm$ and radius of the holes, $a = 200 nm$. The dielectric permittivity of the superstrate is $\varepsilon_1 = 1$, while $\varepsilon_2$ (substrate) is taken for $BaF_2$. The simulations are done for three different frequencies: $861.6 cm^{-1}$, $870.4 cm^{-1}$ and $894 cm^{-1}$. The truncation number, $n_{max}$, used in the main text is highlighted by the vertical black dashed line.

## S4. Comparison between the analytical solution and full-wave simulations for a hole array in α-MoO₃ slab

In this section we present the results of the full-wave simulations for a twisted rectangular HA and compare them with our analytical approximation. Namely, we simulate a HA in α-MoO₃ slab whose crystallographic axes coincide with the cartesian ones. HA is twisted with respect to crystallographic axes by an angle $\theta$, and the incident wave is linearly polarized at an angle $\varphi$ with respect to the *x*-axis (as shown in the inset to Figure S2a). In Figure S2 we represent the reflection p- and s-polarization coefficients, $\delta R_p = \left| \frac{R_{p00} - R_{p00}^b}{R_{p00}^b} \right|$ and $\delta R_s = \left| \frac{R_{s00} - R_{s00}^b}{R_{s00}^b} \right|$, respectively, where the

superscript b stands for the s- and p-polarization reflection coefficient of a bare slab. In Figure S2a, the HA translation vectors coincide with the in-plane axes of α-MoO$_3$, $\theta = 0°$, while the polarization of the incident wave is rotated, $\varphi = 30°$. In this case polarization conversion both for the cases of the bare α-MoO$_3$ slab and HA takes place. In contrast, Figure S2b represents a less symmetric configuration, namely, the case of non-zero polarization angle, $\theta = 30°$ (while the HA is twisted an angle $\varphi = 45°$). The spectra of the relative reflection coefficients represented in Figure S2 are normalized with respect to the ones for the bare slab. Both panels of Figure S2 demonstrate that the analytical solution matches well with the full-wave simulations, both for s- and p-polarizations.

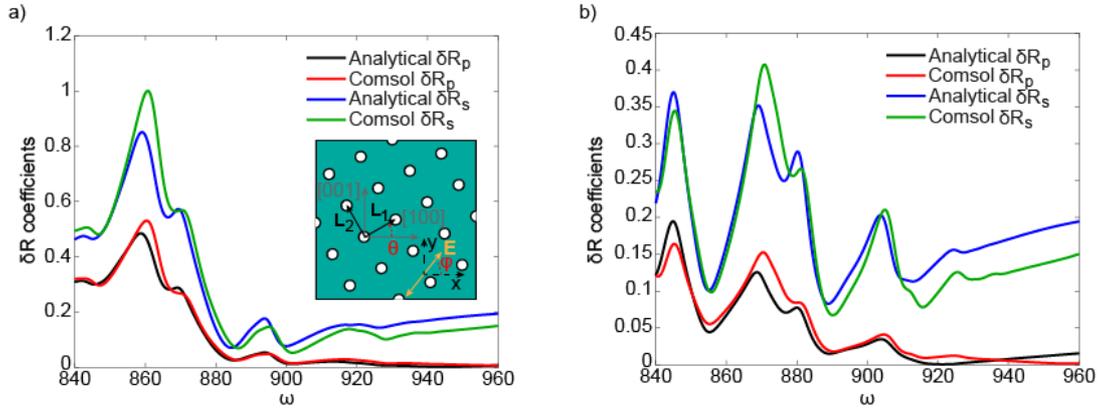

**Figure S2. Comparison between the analytical solution and full-wave simulations for a hole array in α-MoO$_3$ slab. (a)** The spectra of $\delta R_p$ and $\delta R_s$ as a function of the frequency for twisting angles $\theta = 0°$ and $\varphi = 30°$. As input, general case of a twisted square HA with respect to crystallographic axes. Crystallographic axes are aligned with cartesian axes. $\theta$ represents the angle between the HA translation vectors and crystallographic axes is $\theta$, while $\varphi$ is the angle between the incident electric field and x axis. **(b)** The spectra of $\delta R_p$ and $\delta R_s$ as a function of the frequency for $\theta = 30°$ and $\varphi = 45°$.